\documentclass[preprint,superscriptaddress,nofootinbib]{revtex4}
  \usepackage{graphicx}
  \begin{document}
  \title{Study on the ${\Upsilon}(1S)$ ${\to}$ $B_{c}D_{s}$ decay}
  \author{Junfeng Sun}
  \affiliation{Institute of Particle and Nuclear Physics,
              Henan Normal University, Xinxiang 453007, China}
  \author{Yueling Yang}
  \affiliation{Institute of Particle and Nuclear Physics,
              Henan Normal University, Xinxiang 453007, China}
  \author{Qingxia Li}
  \affiliation{Institute of Particle and Nuclear Physics,
              Henan Normal University, Xinxiang 453007, China}
  \author{Haiyan Li}
  \affiliation{Institute of Particle and Nuclear Physics,
              Henan Normal University, Xinxiang 453007, China}
  \author{Na Wang}
  \affiliation{Institute of Particle and Key Laboratory of Quark and Lepton Physics,
              Central China Normal University, Wuhan 430079, China}
  \affiliation{Institute of Particle and Nuclear Physics,
              Henan Normal University, Xinxiang 453007, China}
  \author{Qin Chang}
  \affiliation{Institute of Particle and Nuclear Physics,
              Henan Normal University, Xinxiang 453007, China}
  \author{Jinshu Huang}
  \affiliation{College of Physics and Electronic Engineering,
              Nanyang Normal University, Nanyang 473061, China}
  %%%%%%%%%%%%%%%%%%%%%%%%%%%%%%%%%%%%%%%%%%%%%%%%%%%%%%%%%%%
  \begin{abstract}
  The branching ratio and direct $CP$ asymmetry of the
  ${\Upsilon}(1S)$ ${\to}$ $B_{c}D_{s}$ weak decay
  are estimated with the perturbative QCD approach firstly.
  It is found that (1) The direct $CP$-violating asymmetry is close to zero.
  (2) the branching ratio ${\cal B}r({\Upsilon}(1S){\to}B_{c}D_{s})$
  ${\gtrsim}$ $10^{-10}$ might be measurable
  at the future experiments.
  \end{abstract}
  \pacs{13.25.Gv 12.39.St 14.40.Pq}
  \maketitle

  %%%%%%%%%%%%%%%%%%%%%%%%%%%%%%%%%%%%%%%%%%%%%%%%%%%%%%%%%%%
  \section{Introduction}
  \label{sec01}
  The ${\Upsilon}(1S)$ meson is the ground $S$-wave spin-triplet bottomonium
  (bound state of $b\bar{b}$) with the well-established quantum
  number of $I^{G}J^{PC}$ $=$ $0^{-}1^{--}$ \cite{pdg}.
  Its mass, $m_{{\Upsilon}(1S)}$ $=$ $9460.30{\pm}0.26$ MeV
  \cite{pdg}, is less than the kinematic open-bottom threshold.
  Phenomenologically, the dominated ${\Upsilon}(1S)$ hadronic
  decay through the $b\bar{b}$ pairs annihilation into three gluons,
  with branching ratio ${\cal B}r({\Upsilon}(1S){\to}ggg)$
  $=$ $(81.7{\pm}0.7)\%$ \cite{pdg}, is suppressed by the
  Okubo-Zweig-Iizuka rule \cite{o,z,i}.
  The partial width of the ${\Upsilon}(1S)$ electromagnetic
  decay through the $b\bar{b}$ pairs annihilation into a
  virtual photon, $(3+R){\Gamma}_{{\ell}^{+}{\ell}^{-}}$,
  is proportional to $Q_{b}^{2}$,
  where $Q_{b}$ $=$ $-1/3$ is the electric charge of
  the bottom quark in the unit of ${\vert}e{\vert}$,
  $R$ is the ratio of the inclusive production cross section
  of hadrons to the ${\mu}^{+}{\mu}^{-}$ pair production cross
  section, and ${\Gamma}_{{\ell}^{+}{\ell}^{-}}$ is the partial
  width of the pure leptonic ${\Upsilon}(1S)$ ${\to}$
  ${\ell}^{+}{\ell}^{-}$ decay.
  Besides\footnotemark[1],
  \footnotetext[1]{
  In addition, there are the radiative decay ${\Upsilon}(1S)$
  ${\to}$ ${\gamma}gg$ and the magnetic dipole transition decay
  ${\Upsilon}(1S)$ ${\to}$ ${\gamma}{\eta}_{b}$ \cite{1212.6552}.
  The branching ratio for the radiative decay
  is ${\cal B}r({\Upsilon}(1S){\to}{\gamma}gg)$ $=$ $(2.2{\pm}0.6)\%$
  \cite{pdg}. No signals of the magnetic dipole transition decay
  ${\Upsilon}(1S)$ ${\to}$ ${\gamma}{\eta}_{b}$ have been seen
  experimentally until now.}
  the ${\Upsilon}(1S)$ meson can also decay
  via the weak interactions within the standard model,
  although the branching ratio is very small,
  about $2/{\tau}_{B}{\Gamma}_{{\Upsilon}(1S)}$
  ${\sim}$ ${\cal O}(10^{-8})$ \cite{pdg},
  where ${\tau}_{B}$ and ${\Gamma}_{{\Upsilon}(1S)}$
  are the lifetime of the $B_{u,d,s}$ meson and the
  total width of the ${\Upsilon}(1S)$ meson, respectively.
  In this paper, we will study the ${\Upsilon}(1S)$ ${\to}$
  $B_{c}D_{s}$ weak decays with the perturbative QCD (pQCD)
  approach \cite{pqcd1,pqcd2,pqcd3}.
  The motivation is listed as follows.

  From the experimental point of view,
  (1)
  over $10^{8}$ ${\Upsilon}(1S)$ data samples were
  accumulated by the Belle detector at the KEKB $e^{+}e^{-}$
  asymmetric energy collider \cite{1212.5342}.
  It is hopefully expected that more and more upsilon data
  samples will be collected with great precision at the
  forthcoming SuperKEKB and the running upgraded LHC.
  A large amount of ${\Upsilon}(1S)$ data
  samples offer a realistic possibility to search for
  the ${\Upsilon}(1S)$ weak decays which in some cases
  might be detectable.
  Theoretical studies on the ${\Upsilon}(1S)$
  weak decays are necessary to give a ready
  reference.
  (2)
  For the ${\Upsilon}(1S)$ ${\to}$ $B_{c}D_{s}$ weak decay,
  the back-to-back final states with opposite electric charges
  have definite momentums and energies in the center-of-mass
  frame of the ${\Upsilon}(1S)$ meson.
  In addition, identification of either a single
  flavored $D_{s}$ or $B_{c}$ meson is free from the low
  double-tagging efficiency \cite{zpc62.271},
  and can provide an unambiguous evidence of the
  ${\Upsilon}(1S)$ weak decay.
  Of course, it should be noticed that small branching ratios
  for the ${\Upsilon}(1S)$ weak decays make the observation
  extremely challenging, and any evidences of an abnormally
  large production rate of either a single $D_{s}$ or $B_{c}$
  meson might be a hint of new physics \cite{zpc62.271}.

  From the theoretical point of view,
  the ${\Upsilon}(1S)$ weak decays permit one to crosscheck
  parameters obtained from the $b$-flavored hadron decays,
  to further explore the underlying dynamical mechanism of
  the heavy quark weak decay, and to test various
  phenomenological approaches.
  In recent several years, many attractive methods have been
  developed to evaluate hadronic matrix elements (HME) where
  the local quark-level operators are sandwiched between
  the initial and final hadron states,
  such as pQCD \cite{pqcd1,pqcd2,pqcd3},
  the QCD factorization \cite{qcdf2}
  and the soft and collinear effective theory
  \cite{scet1,scet2,scet3,scet4},
  which could give reasonable explanation for
  many measurements on the nonleptonic $B_{u,d}$ decays.
  The ${\Upsilon}(1S)$ ${\to}$ $B_{c}D_{s}$ weak decay
  is favored by the color factor due to the external
  $W$ emission topological structure, and by the
  Cabibbo-Kobayashi-Maskawa (CKM) factors
  ${\vert}V_{cb}V_{cs}^{\ast}{\vert}$, so it should
  have a large branching ratio.
  However, as far as we know, there is no theoretical
  investigation on the ${\Upsilon}(1S)$ ${\to}$
  $B_{c}D_{s}$ weak decay at the moment.
  In this paper, we will predict the branching ratio
  and direct $CP$-violating asymmetry of the
  ${\Upsilon}(1S)$ ${\to}$ $B_{c}D_{s}$ weak decay
  with the pQCD approach to confirm whether it is
  possible to search for this process at the future
  experiments.

  This paper is organized as follows.
  In section \ref{sec02}, we present the theoretical framework
  and the amplitude for the ${\Upsilon}(1S)$ ${\to}$ $B_{c}D_{s}$
  decay. Section \ref{sec03} is devoted to numerical results
  and discussion. Finally, we conclude with a summary in
  the last section.

  %%%%%%%%%%%%%%%%%%%%%%%%%%%%%%%%%%%%%%%%%%%%%%%%%%%%%%%%%%%
  \section{theoretical framework}
  \label{sec02}
  %%%%%%%%%%%%%%%%%%%%%%%%%%
  \subsection{The effective Hamiltonian}
  \label{sec0201}
  Using the operator product expansion and renormalization
  group equation, the effective Hamiltonian responsible for
  the ${\Upsilon}(1S)$ ${\to}$ $B_{c}D_{s}$ weak decay
  is written as \cite{9512380}
 %---------------------------------------------------------
   \begin{equation}
  {\cal H}_{\rm eff}\, =\,
   \frac{G_{F}}{\sqrt{2}}\,
   \Big\{ V_{cb} V_{cs}^{\ast}
   \sum\limits_{i=1}^{2}
   C_{i}({\mu})\,Q_{i}({\mu})
  -V_{tb} V_{ts}^{\ast}
   \sum\limits_{j=3}^{10}
   C_{j}({\mu})\,Q_{j}({\mu}) \Big\}
   + {\rm H.c.}
   \label{hamilton},
   \end{equation}
 %---------------------------------------------------------
  where $G_{F}$ $=$ $1.166{\times}10^{-5}\,{\rm GeV}^{-2}$ \cite{pdg}
  is the Fermi coupling constant;
  the CKM factors are expressed as a power series in
  the Wolfenstein parameter ${\lambda}$ ${\sim}$ $0.2$ \cite{pdg},
 %---------------------------------------------------------
  \begin{eqnarray}
  V_{cb}V_{cs}^{\ast} &=&
  +            A{\lambda}^{2}
  - \frac{1}{2}A{\lambda}^{4}
  - \frac{1}{8}A{\lambda}^{6}(1+4A^{2})
  +{\cal O}({\lambda}^{8})
  \label{eq:ckm01}, \\
  V_{tb}V_{ts}^{\ast} &=& -V_{cb}V_{cs}^{\ast}
  - A{\lambda}^{4}({\rho}-i{\eta})
  +{\cal O}({\lambda}^{8})
  \label{eq:ckm02}.
  \end{eqnarray}
 %---------------------------------------------------------
  The Wilson coefficients $C_{i}(\mu)$ summarize the
  physical contributions above the scale of ${\mu}$,
  and have been reliably evaluated to the next-to-leading
  logarithmic order.
  The local operators are defined as follows.
 %-----------------------------------------------------
    \begin{eqnarray}
    Q_{1} &=&
  [ \bar{c}_{\alpha}{\gamma}_{\mu}(1-{\gamma}_{5})b_{\alpha} ]
  [ \bar{s}_{\beta} {\gamma}^{\mu}(1-{\gamma}_{5})c_{\beta} ]
    \label{q1}, \\
 %-----------------------------------------------------
    Q_{2} &=&
  [ \bar{c}_{\alpha}{\gamma}_{\mu}(1-{\gamma}_{5})b_{\beta} ]
  [ \bar{s}_{\beta}{\gamma}^{\mu}(1-{\gamma}_{5})c_{\alpha} ]
    \label{q2},
    \end{eqnarray}
 %-----------------------------------------------------
    \begin{eqnarray}
    Q_{3} &=& \sum\limits_{q}
  [ \bar{s}_{\alpha}{\gamma}_{\mu}(1-{\gamma}_{5})b_{\alpha} ]
  [ \bar{q}_{\beta} {\gamma}^{\mu}(1-{\gamma}_{5})q_{\beta} ]
    \label{q3}, \\
 %-----------------------------------------------------
    Q_{4} &=& \sum\limits_{q}
  [ \bar{s}_{\alpha}{\gamma}_{\mu}(1-{\gamma}_{5})b_{\beta} ]
  [ \bar{q}_{\beta}{\gamma}^{\mu}(1-{\gamma}_{5})q_{\alpha} ]
    \label{q4}, \\
 %-----------------------------------------------------
    Q_{5} &=& \sum\limits_{q}
  [ \bar{s}_{\alpha}{\gamma}_{\mu}(1-{\gamma}_{5})b_{\alpha} ]
  [ \bar{q}_{\beta} {\gamma}^{\mu}(1+{\gamma}_{5})q_{\beta} ]
    \label{q5}, \\
 %-----------------------------------------------------
    Q_{6} &=& \sum\limits_{q}
  [ \bar{s}_{\alpha}{\gamma}_{\mu}(1-{\gamma}_{5})b_{\beta} ]
  [ \bar{q}_{\beta}{\gamma}^{\mu}(1+{\gamma}_{5})q_{\alpha} ]
    \label{q6},
    \end{eqnarray}
 %-----------------------------------------------------
    \begin{eqnarray}
    Q_{7} &=& \sum\limits_{q} \frac{3}{2}Q_{q}\,
  [ \bar{s}_{\alpha}{\gamma}_{\mu}(1-{\gamma}_{5})b_{\alpha} ]
  [ \bar{q}_{\beta} {\gamma}^{\mu}(1+{\gamma}_{5})q_{\beta} ]
    \label{q7}, \\
 %-----------------------------------------------------
    Q_{8} &=& \sum\limits_{q} \frac{3}{2}Q_{q}\,
  [ \bar{s}_{\alpha}{\gamma}_{\mu}(1-{\gamma}_{5})b_{\beta} ]
  [ \bar{q}_{\beta}{\gamma}^{\mu}(1+{\gamma}_{5})q_{\alpha} ]
    \label{q8}, \\
 %-----------------------------------------------------
    Q_{9} &=& \sum\limits_{q} \frac{3}{2}Q_{q}\,
  [ \bar{s}_{\alpha}{\gamma}_{\mu}(1-{\gamma}_{5})b_{\alpha} ]
  [ \bar{q}_{\beta} {\gamma}^{\mu}(1-{\gamma}_{5})q_{\beta} ]
    \label{q9}, \\
 %-----------------------------------------------------
    Q_{10} &=& \sum\limits_{q} \frac{3}{2}Q_{q}\,
  [ \bar{s}_{\alpha}{\gamma}_{\mu}(1-{\gamma}_{5})b_{\beta} ]
  [ \bar{q}_{\beta}{\gamma}^{\mu}(1-{\gamma}_{5})q_{\alpha} ]
    \label{q10},
    \end{eqnarray}
 %-----------------------------------------------------
  where $Q_{1,2}$, $Q_{3,{\cdots},6}$, and $Q_{7,{\cdots},10}$
  are usually called as the tree operators, QCD
  penguin operators, and electroweak penguin operators,
  respectively;
  ${\alpha}$ and ${\beta}$ are color indices;
  $q$ denotes all the active quarks at the scale of
  ${\mu}$ ${\sim}$ ${\cal O}(m_{b})$, i.e.,
  $q$ $=$ $u$, $d$, $s$, $c$, $b$.

  %%%%%%%%%%%%%%%%%%%%%%%%%%
  \subsection{Hadronic matrix elements}
  \label{sec0202}
  To obtain the decay amplitudes, the remaining works are
  to calculate the hadronic matrix elements of local
  operators as accurately as possible.
  Based on the $k_{T}$ factorization theorem \cite{npb366}
  and the Lepage-Brodsky approach for exclusive processes \cite{prd22},
  HME can be written as the convolution of hard scattering
  subamplitudes containing perturbative contributions
  with the universal wave functions reflecting the
  nonperturbative contributions with the pQCD approach,
  where the transverse momentums of quarks are retained
  and the Sudakov factors are introduced, in order to regulate
  the endpoint singularities and provide a naturally
  dynamical cutoff on nonperturbative contributions.
  Usually, the decay amplitude can be factorized into three
  parts: the hard effects incorporated into the Wilson
  coefficients $C_{i}$, the process-dependent scattering
  amplitudes $T$, and the
  universal wave functions ${\Phi}$, i.e.,
  %-----------------------------------------------------
  \begin{equation}
  {\int} dx\, db\,
  C_{i}(t)\,T(t,x,b)\,{\Phi}(x,b)e^{-S}
  \label{hadronic},
  \end{equation}
  %-----------------------------------------------------
  where $t$ is a typical scale, $x$ is the
  longitudinal momentum fraction of the valence quark,
  $b$ is the conjugate variable of the transverse
  momentum, and $e^{-S}$ is the Sudakov factor.

  %%%%%%%%%%%%%%%%%%%%%%%%%%
  \subsection{Kinematic variables}
  \label{sec0203}
  The light cone kinematic variables in the ${\Upsilon}(1S)$
  rest frame are defined as follows.
  %------------------------------------
  \begin{eqnarray}
  p_{{\Upsilon}} &=& p_{1}\, =\, \frac{m_{1}}{\sqrt{2}}(1,1,0)
  \label{kine-p1}, \\
  %------------------------------------
  p_{B_{c}} &=& p_{2}\, =\, (p_{2}^{+},p_{2}^{-},0)
  \label{kine-p2}, \\
  %------------------------------------
  p_{D_{s}} &=& p_{3}\, =\, (p_{3}^{-},p_{3}^{+},0)
  \label{kine-p3}, \\
  %------------------------------------
  k_{i} &=& x_{i}\,p_{i}+(0,0,\vec{k}_{iT})
  \label{kine-ki}, \\
  %------------------------------------
  {\epsilon}_{\Upsilon}^{\parallel} &=& \frac{1}{ \sqrt{2} }(1,-1,0)
  \label{kine-1el},
  \end{eqnarray}
  where $x_{i}$ and $\vec{k}_{iT}$ are the
  longitudinal momentum fraction and transverse
  momentum of the valence quark, respectively;
  ${\epsilon}_{\Upsilon}^{\parallel}$ is the
  longitudinal polarization vector of the
  ${\Upsilon}(1S)$ meson.
  The notation of momentum is showed in Fig.\ref{fig1}(a).
  There are some relations among these kinematic variables.
  %------------------------------------
  \begin{eqnarray}
  p_{i}^{\pm} &=& (E_{i}\,{\pm}\,p)/\sqrt{2}
  \label{kine-pipm}, \\
  %------------------------------------
  s &=& 2\,p_{2}{\cdot}p_{3}
  \label{kine-s}, \\
  %------------------------------------
  t &=& 2\,p_{1}{\cdot}p_{2} = 2\,m_{1}\,E_{2}
  \label{kine-t}, \\
  %------------------------------------
  u &=& 2\,p_{1}{\cdot}p_{3} = 2\,m_{1}\,E_{3}
  \label{kine-u},
  \end{eqnarray}
  %------------------------------------
  \begin{equation}
  p = \frac{\sqrt{ [m_{1}^{2}-(m_{2}+m_{3})^{2}]\,[m_{1}^{2}-(m_{2}-m_{3})^{2}] }}{2\,m_{1}}
  \label{kine-pcm},
  \end{equation}
  %------------------------------------
  where $p$ is the common momentum of
  the final $B_{c}$ and $D_{s}$ states;
  $m_{1}$ $=$ $m_{{\Upsilon}(1S)}$,
  $m_{2}$ $=$ $m_{B_{c}}$ and
  $m_{3}$ $=$ $m_{D_{s}}$ denote the masses
  of the ${\Upsilon}(1S)$, $B_{c}$ and $D_{s}$
  mesons, respectively.

  %%%%%%%%%%%%%%%%%%%%%%%%%%
  \subsection{Wave functions}
  \label{sec0204}
  The HME of diquark operators squeezed
  between the vacuum and ${\Upsilon}(1S)$, $B_{c}$, $D_{s}$
  mesons are defined as follows.
  %------------------------------------
  \begin{equation}
 {\langle}0{\vert}b_{i}(z)\bar{b}_{j}(0){\vert}
 {\Upsilon}(p_{1},{\epsilon}_{\parallel}){\rangle}\,
 =\, \frac{1}{4}f_{\Upsilon}
 {\int}dk_{1}\,e^{-ik_{1}{\cdot}z}
  \Big\{ \!\!\not{\epsilon}_{\parallel} \Big[
   m_{1}\,{\phi}_{\Upsilon}^{v}(k_{1})
  -\!\!\not{p}_{1}\, {\phi}_{\Upsilon}^{t}(k_{1})
  \Big] \Big\}_{ji}
  \label{wave-bbl},
  \end{equation}
  %------------------------------------
  \begin{equation}
 {\langle}B_{c}^{+}(p_{2}){\vert}\bar{c}_{i}(z)b_{j}(0){\vert}0{\rangle}\,
 =\, \frac{i}{4}f_{B_{c}} {\int}dk_{2}\,e^{ik_{2}{\cdot}z}\,
  \Big\{ {\gamma}_{5}\Big[ \!\!\not{p}_{2}+m_{2}\Big]
 {\phi}_{B_{c}}(k_{2}) \Big\}_{ji}
  \label{wave-bcp},
  \end{equation}
  %------------------------------------
  \begin{equation}
 {\langle}D_{s}^{-}(p_{3}){\vert}\bar{s}_{i}(z)c_{j}(0)
 {\vert}0{\rangle}\ =\
  \frac{i}{4}f_{D_{s}} {\int}_{0}^{1}dk_{3}\,e^{ik_{3}{\cdot}z}
  \Big\{ {\gamma}_{5}\Big[ \!\!\not{p}_{3}+m_{3}\Big]
 {\Phi}_{D_{s}}(k_{3}) \Big\}_{ji}
  \label{wave-ds},
  \end{equation}
  %------------------------------------
  where $f_{\Upsilon}$, $f_{B_{c}}$, $f_{D_{s}}$ are
  decay constants.

  There are several phenomenological models for the $D_{s}$
  meson wave functions (for example, Eq.(30) in Ref.\cite{prd78lv}).
  In this paper, we will take the model favored by
  Ref.\cite{prd78lv} via fitting with measurements on
  the $B$ ${\to}$ $DP$ decays.
  %-----------------------------------------------------
   \begin{equation}
  {\phi}_{D_{s}}(x,b)=6\,x\bar{x}\,
   \Big\{ 1+ C_{D}(1-2\,x) \Big\}\,
  {\exp}\Big\{ -\frac{1}{2}\,w^2\,b^2 \Big\}
   \label{DA-ds},
   \end{equation}
 %-----------------------------------------------------
  where $\bar{x}$ $=$ $1$ $-$ $x$;
  $x$ and $b$ are the longitudinal momentum fraction
  and the conjugate variable of the transverse momentum $k_{T}$
  of the strange quark in the $D_{s}$ meson, respectively;
  the exponential term represents the $k_{T}$ distribution;
  $C_{D}$ $=$ $0.4{\pm}0.1$ and $w$ $=$ $0.2$ GeV \cite{prd78lv}.

  Due to $m_{{\Upsilon}(1S)}$ ${\simeq}$ $2m_{b}$
  and $m_{B_{c}}$ ${\simeq}$ $m_{b}$ $+$ $m_{c}$,
  nonrelativistic quantum chromodynamics
  \cite{prd46,prd51,rmp77} and
  Schr\"{o}dinger equation can be used to describe
  both ${\Upsilon}(1S)$ and $B_{c}$ mesons.
  The wave functions of an isotropic harmonic
  oscillator potential are given in Ref. \cite{prd92},
  %-----------------------------------------------------
   \begin{equation}
  {\phi}_{\Upsilon}^{v}(x) = A\, x\bar{x}\,
  {\exp}\Big\{ -\frac{m_{b}^{2}}{8\,{\beta}_{1}^{2}\,x\,\bar{x}} \Big\}
   \label{wave-bbv},
   \end{equation}
  %-----------------------------------------------------
   \begin{equation}
  {\phi}_{\Upsilon}^{t}(x) = B\, (x-\bar{x})^{2}\,
  {\exp}\Big\{ -\frac{m_{b}^{2}}{8\,{\beta}_{1}^{2}\,x\,\bar{x}} \Big\}
   \label{wave-bbt},
   \end{equation}
  %-----------------------------------------------------
   \begin{equation}
  {\phi}_{B_{c}}(x) = C\, x\bar{x}\,
  {\exp}\Big\{ -\frac{\bar{x}\,m_{c}^{2}+x\,m_{b}^{2}}
                     {8\,{\beta}_{2}^{2}\,x\,\bar{x}} \Big\}
   \label{wave-bc},
   \end{equation}
  %-----------------------------------------------------
   where ${\beta}_{i}$ $=$ ${\xi}_{i}{\alpha}_{s}({\xi}_{i})$
   with ${\xi}_{i}$ $=$ $m_{i}/2$;
   parameters $A$, $B$, $C$ are the normalization coefficients
   satisfying the following conditions
  %-----------------------------------------------------
   \begin{equation}
  {\int}_{0}^{1}dx\,{\phi}_{\Upsilon}^{v,t}(x) =1,
   \quad
  {\int}_{0}^{1}dx\,{\phi}_{B_{c}}(x)=1.
   \label{wave-abc}.
   \end{equation}
  %-----------------------------------------------------

  %%%%%%%%%%%%%%%%%%%%%%%%%%
  \subsection{Decay amplitudes}
  \label{sec0205}
\begin{figure}[h]
  \includegraphics[width=0.95\textwidth,bb=80 530 530 720]{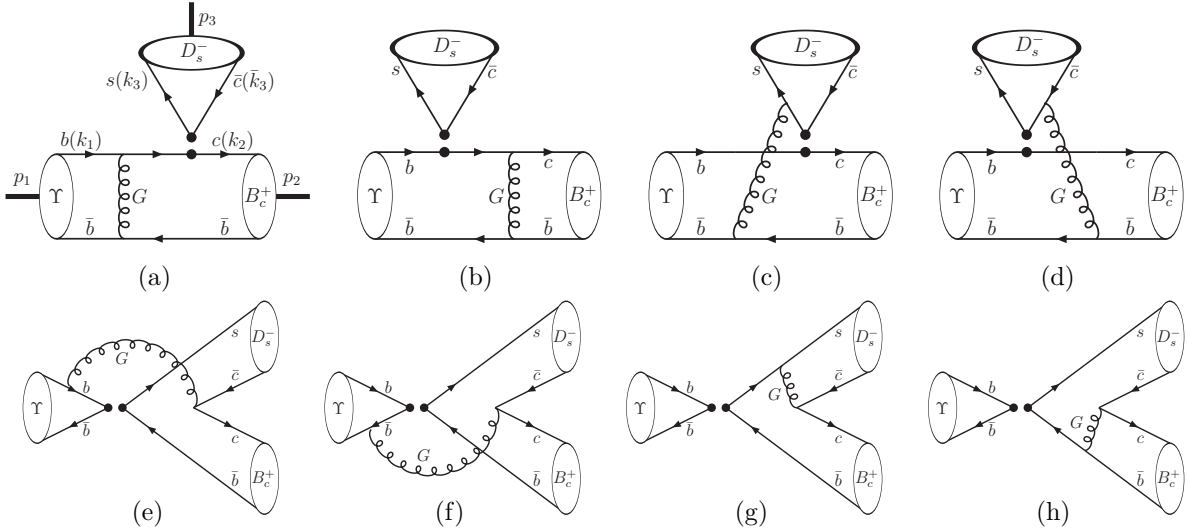}
  \caption{Feynman diagrams for the ${\Upsilon}(1S)$ ${\to}$
  $B_{c}D_{s}$ decay with the pQCD approach, including
  the factorizable emission diagrams (a,b),
  the nonfactorizable emission diagrams (c,d),
  the nonfactorizable annihilation diagrams (e,f),
  and the factorizable annihilation diagrams (g,h).}
  \label{fig1}
  \end{figure}

  The Feynman diagrams for the ${\Upsilon}(1S)$ ${\to}$
  $B_{c}D_{s}$ decay are shown in Fig.\ref{fig1}.
  There are two types: the emission and annihilation
  topologies, where diagrams containing gluon exchanges
  between the quarks in the same (different) mesons
  are entitled (non)factorizable diagrams.
  %-----------------------------------------------------
  %-----------------------------------------------------

  By calculating these diagrams with the pQCD master
  formula Eq.(\ref{hadronic}), the decay amplitudes
  of ${\Upsilon}(1S)$ ${\to}$ $B_{c}D_{s}$
  decay can be expressed as:
  %-----------------------------------------------------
   \begin{eqnarray}
   \lefteqn{ {\cal A}({\Upsilon}(1S){\to}B_{c}D_{s})\ =\
   \sqrt{2}\,G_{F}\,{\pi}\,f_{\Upsilon}\,f_{B_{c}}\,f_{D_{s}}\,
   \frac{C_{F}}{N}\,m_{{\Upsilon}}^{3}\,({\epsilon}_{\Upsilon}{\cdot}p_{D_{s}}) }
   \nonumber \\ &{\times}&
   \Big\{ V_{cb} V_{cs}^{\ast}\,\Big[ {\cal A}_{a+b}^{LL}\,a_{1}
   +{\cal A}_{c+d}^{LL}\,C_{2} \Big]
   - V_{tb} V_{ts}^{\ast}\,\Big[ {\cal A}_{a+b}^{LL}\,(a_{4}+a_{10})
   \nonumber \\ & &
   + {\cal A}_{a+b}^{SP}\, (a_{6}+a_{8})
   + {\cal A}_{c+d}^{LL}\, (C_{3}+C_{9})
   + {\cal A}_{c+d}^{SP}\, (C_{5}+C_{7})
   \nonumber \\ & &
   + {\cal A}_{e+f}^{LL}\, (C_{3}+C_{4}-\frac{1}{2}C_{9}-\frac{1}{2}C_{10})
   + {\cal A}_{e+f}^{LR}\, (C_{6}-\frac{1}{2}C_{8})
   \nonumber \\ & &
   + {\cal A}_{g+h}^{LL}\, (a_{3}+a_{4}-\frac{1}{2}a_{9}-\frac{1}{2}a_{10})
   + {\cal A}_{g+h}^{LR}\, (a_{5}-\frac{1}{2}a_{7})
   \nonumber \\ & &
   + {\cal A}_{e+f}^{SP}\, (C_{5}-\frac{1}{2}C_{7}) \Big] \Big\}
   \label{amp-all},
   \end{eqnarray}
  %-----------------------------------------------------
  where $C_{F}$ $=$ $4/3$ and the color number $N$ $=$ $3$.

  The parameters $a_{i}$ are defined as follows.
  %-----------------------------------------------------
  \begin{eqnarray}
  a_{i} &=& C_{i}+C_{i+1}/N, \quad (i=1,3,5,7,9);
  \label{eq:ai1357} \\
  a_{i} &=& C_{i}+C_{i-1}/N, \quad (i=2,4,5,6,10).
  \label{eq:ai2468}
  \end{eqnarray}
  %-----------------------------------------------------

  The building blocks ${\cal A}_{a+b}$, ${\cal A}_{c+d}$,
  ${\cal A}_{e+f}$, ${\cal A}_{g+h}$ denote the contributions of
  the factorizable emission diagrams Fig.\ref{fig1}(a,b),
  the nonfactorizable emission diagrams Fig.\ref{fig1}(c,d),
  the nonfactorizable annihilation diagrams Fig.\ref{fig1}(e,f),
  the factorizable annihilation diagrams Fig.\ref{fig1}(g,h),
  respectively. They are defined as
  \begin{equation}
  {\cal A}_{i+j}^{k} = {\cal A}_{i}^{k}+{\cal A}_{j}^{k}
  \label{eq:ampij},
  \end{equation}
  where the subscripts $i$ and $j$ correspond
  to the indices of Fig.\ref{fig1};
  the superscript $k$ refers
  to one of the three possible Dirac structures, namely
  $k$ $=$ $LL$ for $(V-A){\otimes}(V-A)$,
  $k$ $=$ $LR$ for $(V-A){\otimes}(V+A)$, and
  $k$ $=$ $SP$ for $-2(S-P){\otimes}(S+P)$.
  The explicit expressions of these building blocks
  are collected in the Appendix \ref{blocks}.

  %%%%%%%%%%%%%%%%%%%%%%%%%%%%%%%%%%%%%%%%%%%%%%%%%%%%%%%%%%%
  \section{Numerical results and discussion}
  \label{sec03}
  In the rest frame of the ${\Upsilon}(1S)$ meson,
  the $CP$-averaged branching ratio and direct $CP$-violating
  asymmetry for the ${\Upsilon}(1S)$ ${\to}$
  $B_{c}D_{s}$ weak decay are written as
 %-----------------------------------------------------
   \begin{equation}
  {\cal B}r({\Upsilon}(1S){\to}B_{c}D_{s})\ =\ \frac{1}{12{\pi}}\,
   \frac{p}{m_{{\Upsilon}}^{2}{\Gamma}_{{\Upsilon}}}\,
  {\vert}{\cal A}({\Upsilon}(1S){\to}B_{c}D_{s}){\vert}^{2}
   \label{br},
   \end{equation}
 %-----------------------------------------------------
   \begin{equation}
  {\cal A}_{\rm CP}({\Upsilon}(1S){\to}B_{c}D_{s})\ =\
   \frac{{\cal B}r({\Upsilon}(1S){\to}B_{c}^{+}D_{s}^{-})
        -{\cal B}r({\Upsilon}(1S){\to}B_{c}^{-}D_{s}^{+})}
        {{\cal B}r({\Upsilon}(1S){\to}B_{c}^{+}D_{s}^{-})
        +{\cal B}r({\Upsilon}(1S){\to}B_{c}^{-}D_{s}^{+})}
   \label{cp},
   \end{equation}
 %-----------------------------------------------------
  where the decay width ${\Gamma}_{\Upsilon}$ $=$
  $54.02{\pm}1.25$ keV \cite{pdg}.

  The numerical values of other input parameters are listed as follows.

  (1) The Wolfenstein parameters \cite{pdg}:
    $A$ $=$ $0.814^{+0.023}_{-0.024}$,
    ${\lambda}$ $=$ $0.22537{\pm}0.00061$,
    $\bar{\rho}$ $=$ $0.117{\pm}0.021$, and
    $\bar{\eta}$ $=$ $0.353{\pm}0.013$, where
    $({\rho}+i{\eta})$ $=$ $(\bar{\rho}+i\bar{\eta})
    (1+{\lambda}^{2}/2+{\cdots})$.

  (2) Masses of quarks \cite{pdg}:
    $m_{c}$ $=$ $1.67{\pm}0.07$ GeV and
    $m_{b}$ $=$ $4.78{\pm}0.06$ GeV.

  (3) Decay constants:
    $f_{{\Upsilon}(1S)}$ $=$ $676.4{\pm}10.7$ MeV \cite{prd92},
    $f_{B_{c}}$ $=$ $489{\pm}5$ MeV \cite{fbc}, and
    $f_{D_{s}}$ $=$ $257.5{\pm}4.6$ MeV \cite{pdg}.

  Finally, we get
  %-----------------------------------------------------
   \begin{equation}
  {\cal B}r({\Upsilon}(1S){\to}B_{c}D_{s})\ =\
  (3.78^{+  0.27+  0.42+  0.50+  0.34}_{ -0.26 -0.38 -0.25 -0.32}){\times}10^{-10}
   \label{vbr},
   \end{equation}
 %-----------------------------------------------------
   \begin{equation}
  {\cal A}_{\rm CP}({\Upsilon}(1S){\to}B_{c}D_{s})\ =\
  (4.79^{+  0.21+  1.14+  0.18+  0.36}_{ -0.20 -1.00 -0.44 -0.39}){\times}10^{-5}
   \label{vcp},
   \end{equation}
 %-----------------------------------------------------
  where the central values are obtained with the
  central values of input parameters;
  the first uncertainties come from the CKM parameters;
  the second uncertainties are due to the variation of
  mass $m_{b}$ and $m_{c}$;
  the third uncertainties arise from the typical scale
  ${\mu}$ $=$ $(1{\pm}0.1)t_{i}$, where the expressions
  of $t_{i}$ for different topologies are given in
  Eqs.(\ref{tab}-\ref{tgh});
  and the fourth uncertainties correspond to the variation of
  decay constants $f_{\Upsilon}$, $f_{B_{c}}$, $f_{D_{s}}$
  and shape parameter $C_{D}$ in Eq.(\ref{DA-ds}).
  There are some comments.

  (1)
  It is seen from Eq.(\ref{vbr}) that branching ratio for
  the ${\Upsilon}(1S)$ ${\to}$ $B_{c}D_{s}$ decay can
  reach up to $10^{-10}$, which might be accessible at the
  running LHC and forthcoming SuperKEKB.
  For example, the ${\Upsilon}(1S)$ production cross
  section in p-Pb collision is a few ${\mu}b$
  with the LHCb \cite{jhep1407} and ALICE \cite{plb740}
  detectors at LHC.
  Over $10^{12}$ ${\Upsilon}(1S)$ mesons per $ab^{-1}$
  data collected at LHCb and ALICE  are in principle available,
  corresponding to a few hundreds of the ${\Upsilon}(1S)$ ${\to}$
  $B_{c}D_{s}$ events.

  (2)
  Compared the ${\Upsilon}(1S)$ ${\to}$ $B_{c}D_{s}$ decay
  with the ${\Upsilon}(1S)$ ${\to}$ $B_{c}{\pi}$ decay \cite{prd92},
  they are both the color-favored and CKM-favored.
  There is only the emission topologies and only the
  tree operators contributing to the ${\Upsilon}(1S)$
  ${\to}$ $B_{c}{\pi}$ decay.
  Besides the emission topologies and tree operators,
  there are other contributions from the annihilation
  topologies and penguin operators for the ${\Upsilon}(1S)$
  ${\to}$ $B_{c}D_{s}$ decay.
  In addition, there is another important factor,
  the decay constant $f_{D_{s}}$ $>$ $2f_{\pi}$.
  This might explain the fact that
  although the final phase
  spaces for the ${\Upsilon}(1S)$ ${\to}$ $B_{c}D_{s}$ decay
  is more compact than those for the ${\Upsilon}(1S)$
  ${\to}$ $B_{c}{\pi}$ decay, there is still the relation\footnotemark[2],
  \footnotetext[2]{The branching ratio for the
  ${\Upsilon}(1S)$ ${\to}$ $B_{c}{\pi}$ decay is
  about ${\cal B}r({\Upsilon}(1S){\to}B_{c}{\pi})$
  ${\sim}$ ${\cal O}(10^{-11})$ \cite{prd92}
  with the pQCD approach.}
  ${\cal B}r({\Upsilon}(1S){\to}B_{c}D_{s})$
  $>$ ${\cal B}r({\Upsilon}(1S){\to}B_{c}{\pi})$
  with the pQCD approach.

  (3)
  It is shown from Eq.(\ref{vcp}) that the direct $CP$ asymmetry
  for the ${\Upsilon}(1S)$ ${\to}$ $B_{c}D_{s}$ decay is close
  to zero. The fact should be so.
  As it is well known, the magnitude of direct $CP$
  asymmetry is proportional to the sine of weak phase difference.
  First and foremost, the weak phase difference between the CKM factors
  $V_{cb}V_{cs}^{\ast}$ and $V_{tb}V_{ts}^{\ast}$ are
  suppressed by the factor of ${\lambda}^{2}$.
  Secondly, compared with the tree contributions
  appearing with $V_{cb}V_{cs}^{\ast}$,
  the penguin and annihilation contributions always
  accompanied with $V_{tb}V_{ts}^{\ast}$
  are suppressed by the small Wilson coefficients.

\begin{figure}[h]
  \includegraphics[width=0.5\textwidth]{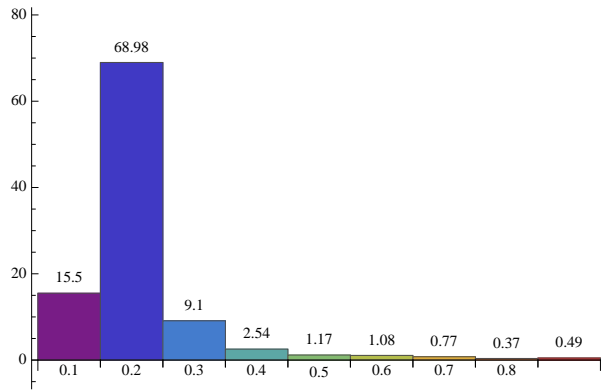}
  \caption{The contributions to the branching ratio from different
  region of ${\alpha}_{s}/{\pi}$ (horizontal axises), where the
  numbers over histogram denote the percentage of the corresponding
  contributions.}
  \label{fig:br-as}
  \end{figure}

  (4)
  As it is well known, due to mass $m_{B_{c}}$ $>$ $m_{{\Upsilon}(1S)}/2$,
  the momentum transition in the ${\Upsilon}(1S)$ ${\to}$ $B_{c}D_{s}$
  decay may be not large enough. One might question whether the
  pQCD approach is applicable and whether the perturbative calculation
  is reliable. Therefore, it is necessary to check what percentage
  of the contributions comes from the perturbative region.
  The contributions to branching ratio from different region of
  ${\alpha}_{s}/{\pi}$ are showed in Fig.(\ref{fig:br-as}).
  One can clearly see from Fig.(\ref{fig:br-as}) that
  more than 90\% contributions to branching ratio come from
  the ${\alpha}_{s}/{\pi}$ ${\le}$ $0.3$ region, and the
  contributions from nonperturbative region with large
  ${\alpha}_{s}/{\pi}$ are highly suppressed. One important
  reason is that assisting with the typical
  scale in Eqs.(\ref{tab}-\ref{tgh}), the quark transverse momentum is
  retained and the Sudakov factor is introduced to effectively
  suppress the nonperturbative contributions within
  the pQCD approach \cite{pqcd1,pqcd2,pqcd3}.

  %-----------------------------------------------------
  %-----------------------------------------------------

  (5)
  There are many uncertainties on our results.
  Other factors, such as the contributions of higher
  order corrections to HME, relativistic effects
  and so on, which are not considered here,
  deserve the dedicated study.
  Our results just provide an order of magnitude estimation.

  %%%%%%%%%%%%%%%%%%%%%%%%%%%%%%%%%%%%%%%%%%%%%%%%%%%%%%%%%%%
  \section{Summary}
  \label{sec04}
  The ${\Upsilon}(1S)$ weak decay is legal within
  the standard model.
  With the potential prospects of the ${\Upsilon}(1S)$
  at high-luminosity dedicated heavy-flavor factories,
  the ${\Upsilon}(1S)$ ${\to}$ $B_{c}D_{s}$,
  weak decays are studied with the pQCD approach.
  It is found that with the nonrelativistic wave functions
  for ${\Upsilon}(1S)$ and $B_{c}$ mesons, branching
  ratios ${\cal B}r({\Upsilon}(1S){\to}B_{c}D_{s})$
  ${\gtrsim}$ $10^{-10}$, which might be
  measurable in future experiments. The direct $CP$-violating
  asymmetry for the ${\Upsilon}(1S)$ ${\to}$ $B_{c}D_{s}$
  decay is close to zero because of the tiny weak phase difference.

  %%%%%%%%%%%%%%%%%%%%%%%%%%%%%%%%%%%%%%%%%%%%%%%%%%%%%%%%%%%
  \section*{Acknowledgments}
  We thank Professor Dongsheng Du (IHEP@CAS) and Professor
  Yadong Yang (CCNU) for helpful discussion.
  We thank the referees for their constructive suggestions.

  \begin{appendix}
  %%%%%%%%%%%%%%%%%%%%%%%%%%%%%%%%%%%%%%%%%%%%%%%%%%%%%%%%%%%
  \section{The building blocks of decay amplitudes}
  \label{blocks}
  For the sake of simplicity,
  we decompose the decay amplitude Eq.(\ref{amp-all})
  into some building blocks ${\cal A}_{i}^{k}$, where
  the subscript $i$ on ${\cal A}_{i}^{k}$ corresponds to
  the indices of Fig.\ref{fig1};
  the superscript $k$ on ${\cal A}_{i}^{k}$ refers
  to one of the three possible Dirac structures
  ${\Gamma}_{1}{\otimes}{\Gamma}_{2}$ of the
  four-quark operator
  $(\bar{q}_{1}{\Gamma}_{1}q_{2})(\bar{q}_{1}{\Gamma}_{2}q_{2})$,
  namely
  $k$ $=$ $LL$ for $(V-A){\otimes}(V-A)$,
  $k$ $=$ $LR$ for $(V-A){\otimes}(V+A)$, and
  $k$ $=$ $SP$ for $-2(S-P){\otimes}(S+P)$.
  The explicit expressions of ${\cal A}_{i}^{k}$
  are written as follows.
  %-----------------------------------------------------
   \begin{eqnarray}
  {\cal A}_{a}^{LL} &=&
  {\int}_{0}^{1}dx_{1} {\int}_{0}^{1}dx_{2}
  {\int}_{0}^{\infty}b_{1} db_{1}
  {\int}_{0}^{\infty}b_{2} db_{2}\,
  {\phi}_{\Upsilon}^{v}(x_{1})\,
  {\phi}_{B_{c}}(x_{2})
   \nonumber \\ & &
  E_{a}(t_{a})\, {\alpha}_{s}(t_{a})\,
  H_{ab}({\alpha}_{e},{\beta}_{a},b_{1},b_{2})\,
   \Big\{ x_{2}+r_{3}^{2}\bar{x}_{2}
   + r_{2}r_{b} \Big\}
   \label{amp-figa-01}, \\
  %-----------------------------------------------------
  {\cal A}_{a}^{SP} &=&
   -2\,r_{3}\,{\int}_{0}^{1}dx_{1} {\int}_{0}^{1}dx_{2}
  {\int}_{0}^{\infty}b_{1} db_{1}
  {\int}_{0}^{\infty}b_{2} db_{2}\,
  {\phi}_{\Upsilon}^{v}(x_{1})\,
  {\phi}_{B_{c}}(x_{2})
   \nonumber \\ & &
  E_{a}(t_{a})\, {\alpha}_{s}(t_{a})\,
  H_{ab}({\alpha}_{e},{\beta}_{a},b_{1},b_{2})\,
   \Big\{ r_{b}+r_{2}\bar{x}_{2} \Big\}
   \label{amp-figa-03},
   \end{eqnarray}
  %-----------------------------------------------------
  %-----------------------------------------------------
   \begin{eqnarray}
  {\cal A}_{b}^{LL} &=&
  {\int}_{0}^{1}dx_{1} {\int}_{0}^{1}dx_{2}
  {\int}_{0}^{\infty}b_{1} db_{1}
  {\int}_{0}^{\infty}b_{2} db_{2}\,
  {\phi}_{B_{c}}(x_{2})\, E_{b}(t_{b})\,
  {\alpha}_{s}(t_{b})
   \nonumber \\ & &
  H_{ab}({\alpha}_{e},{\beta}_{b},b_{2},b_{1})\,
   \Big\{ {\phi}_{\Upsilon}^{v}(x_{1}) \Big[
   2\,r_{2}\,r_{c}-r_{2}^{2}\,x_{1}-r_{3}^{2}\,\bar{x}_{1} \Big]
   \nonumber \\ & &
   + {\phi}_{\Upsilon}^{t}(x_{1}) \Big[ 2\,r_{2}\,x_{1}-r_{c} \Big] \Big\}
   \label{amp-figb-01}, \\
  %-----------------------------------------------------
  {\cal A}_{b}^{SP} &=& -2\,r_{3}\,
  {\int}_{0}^{1}dx_{1} {\int}_{0}^{1}dx_{2}
  {\int}_{0}^{\infty}b_{1} db_{1}
  {\int}_{0}^{\infty}b_{2} db_{2}\,
  {\phi}_{B_{c}}(x_{2})\, E_{b}(t_{b})\,
  {\alpha}_{s}(t_{b})
   \nonumber \\ & &
  H_{ab}({\alpha}_{e},{\beta}_{b},b_{2},b_{1})\,
   \Big\{ {\phi}_{\Upsilon}^{v}(x_{1})\,( 2\,r_{2}-r_{c})
 -{\phi}_{\Upsilon}^{t}(x_{1})\,\bar{x}_{1}\Big\}
   \label{amp-figb-03},
   \end{eqnarray}
  %-----------------------------------------------------
  %-----------------------------------------------------
   \begin{eqnarray}
  {\cal A}_{c}^{LL} &=& \frac{1}{N}
  {\int}_{0}^{1}dx_{1} {\int}_{0}^{1}dx_{2} {\int}_{0}^{1}dx_{3}
  {\int}_{0}^{\infty} db_{1}
  {\int}_{0}^{\infty}b_{2} db_{2}
  {\int}_{0}^{\infty}b_{3} db_{3}\,
  {\delta}(b_{1}-b_{2})
   \nonumber \\ & &
  {\phi}_{B_{c}}(x_{2})\, {\phi}_{D_{s}}(x_{3},b_{3})\,
  E_{c}(t_{c})\,{\alpha}_{s}(t_{c})\,
  H_{cd}({\alpha}_{e},{\beta}_{c},b_{2},b_{3})\,
   \Big\{ {\phi}_{\Upsilon}^{v}(x_{1})
   \nonumber \\ & &
   \Big[ \frac{s\,(x_{1}-\bar{x}_{3}) }{m_{1}^{2}}\,
   +2\,r_{2}^{2}\,(x_{1}-x_{2}) \Big]
   +{\phi}_{\Upsilon}^{t}(x_{1})\, r_{2}\,(x_{2}-x_{1}) \Big\}
   \label{amp-figc-01}, \\
  %-----------------------------------------------------
  {\cal A}_{c}^{SP} &=& -\frac{1}{N}\,r_{3}
  {\int}_{0}^{1}dx_{1} {\int}_{0}^{1}dx_{2} {\int}_{0}^{1}dx_{3}
  {\int}_{0}^{\infty} db_{1}
  {\int}_{0}^{\infty}b_{2} db_{2}
  {\int}_{0}^{\infty}b_{3} db_{3}\,{\alpha}_{s}(t_{c})
   \nonumber \\ & &
  {\delta}(b_{1}-b_{2})\,
  {\phi}_{B_{c}}(x_{2})\, {\phi}_{D_{s}}(x_{3},b_{3})\,
  E_{c}(t_{c})\,
  H_{cd}({\alpha}_{e},{\beta}_{c},b_{2},b_{3})
   \nonumber \\ & &
   \Big\{ {\phi}_{\Upsilon}^{v}(x_{1})\,r_{2}\,
   (\bar{x}_{3}-x_{2})
   +{\phi}_{\Upsilon}^{t}(x_{1})\, (x_{1}-\bar{x}_{3}) \Big\}
   \label{amp-figc-03},
   \end{eqnarray}
  %-----------------------------------------------------
  %-----------------------------------------------------
   \begin{eqnarray}
  {\cal A}_{d}^{LL} &=& \frac{1}{N}
  {\int}_{0}^{1}dx_{1} {\int}_{0}^{1}dx_{2} {\int}_{0}^{1}dx_{3}
  {\int}_{0}^{\infty} db_{1}
  {\int}_{0}^{\infty}b_{2} db_{2}
  {\int}_{0}^{\infty}b_{3} db_{3}\,
  {\delta}(b_{1}-b_{2})
   \nonumber \\ & &
  {\phi}_{B_{c}}(x_{2})\, {\phi}_{D_{s}}(x_{3},b_{3})\,
  E_{d}(t_{d})\,{\alpha}_{s}(t_{d})\,
  H_{cd}({\alpha}_{e},{\beta}_{d},b_{2},b_{3})\,
   \Big\{ {\phi}_{\Upsilon}^{v}(x_{1})
   \nonumber \\ & &
   \Big[ \frac{s\,(x_{3}-x_{2})}{m_{1}^{2}}-r_{3}\,r_{c} \Big]
   +{\phi}_{\Upsilon}^{t}(x_{1})\,r_{2}\,(x_{2}-x_{1}) \Big\}
   \label{amp-figd-01}, \\
  %-----------------------------------------------------
  {\cal A}_{d}^{SP} &=& -\frac{1}{N}\,r_{3}
  {\int}_{0}^{1}dx_{1} {\int}_{0}^{1}dx_{2} {\int}_{0}^{1}dx_{3}
  {\int}_{0}^{\infty} db_{1}
  {\int}_{0}^{\infty}b_{2} db_{2}
  {\int}_{0}^{\infty}b_{3} db_{3}\,{\alpha}_{s}(t_{d})
   \nonumber \\ & &
  {\delta}(b_{1}-b_{2})\,
  {\phi}_{B_{c}}(x_{2})\, {\phi}_{D_{s}}(x_{3},b_{3})\,
  E_{d}(t_{d})\,
  H_{cd}({\alpha}_{e},{\beta}_{d},b_{2},b_{3})
   \nonumber \\ & &
   \Big\{ {\phi}_{\Upsilon}^{v}(x_{1})\, r_{2}\, (r_{c}/r_{3}+x_{2}-x_{3})
 +{\phi}_{\Upsilon}^{t}(x_{1})\,(x_{3}-x_{1}-r_{c}/r_{3}) \Big\}
   \label{amp-figd-03},
   \end{eqnarray}
  %-----------------------------------------------------
  %-----------------------------------------------------
   \begin{eqnarray}
  {\cal A}_{e}^{LL} &=& \frac{1}{N}
  {\int}_{0}^{1}dx_{1} {\int}_{0}^{1}dx_{2} {\int}_{0}^{1}dx_{3}
  {\int}_{0}^{\infty}b_{1} db_{1} {\int}_{0}^{\infty}b_{2} db_{2}
  {\int}_{0}^{\infty} db_{3}\, {\delta}(b_{2}-b_{3})
   \nonumber \\ & &
  {\phi}_{B_{c}}(x_{2})\, {\phi}_{D_{s}}(x_{3},b_{3})\,
  E_{e}(t_{e})\,{\alpha}_{s}(t_{e})\,
  H_{ef}({\alpha}_{a},{\beta}_{e},b_{1},b_{2})\,
   \Big\{ {\phi}_{\Upsilon}^{v}(x_{1})
   \nonumber \\ & &
   \Big[ \frac{s\,(x_{1}-\bar{x}_{3})}{m_{1}^{2}}
   +2\,r_{2}^{2}\,(x_{1}-x_{2})
   +r_{2}\,r_{3}\,(x_{2}-\bar{x}_{3}) \Big]
   -r_{b}\,{\phi}_{\Upsilon}^{t}(x_{1}) \Big\}
   \label{amp-fige-01}, \\
  %-----------------------------------------------------
  {\cal A}_{e}^{LR} &=& \frac{1}{N}
  {\int}_{0}^{1}dx_{1} {\int}_{0}^{1}dx_{2} {\int}_{0}^{1}dx_{3}
  {\int}_{0}^{\infty}b_{1} db_{1} {\int}_{0}^{\infty}b_{2} db_{2}
  {\int}_{0}^{\infty} db_{3}\, {\delta}(b_{2}-b_{3})
   \nonumber \\ & &
  {\phi}_{B_{c}}(x_{2})\, {\phi}_{D_{s}}(x_{3},b_{3})\,
  E_{e}(t_{e})\,{\alpha}_{s}(t_{e})\,
  H_{ef}({\alpha}_{a},{\beta}_{e},b_{1},b_{2})\,
   \Big\{ {\phi}_{\Upsilon}^{v}(x_{1})
   \nonumber \\ & &
   \Big[ \frac{s\,(x_{2}-x_{1})}{m_{1}^{2}}
   +2\,r_{3}^{2}\, (\bar{x}_{3}-x_{1})
   +r_{2}\,r_{3}\, (x_{2}-\bar{x}_{3}) \Big]
   +r_{b}\,{\phi}_{\Upsilon}^{t}(x_{1}) \Big\}
   \label{amp-fige-02}, \\
  %-----------------------------------------------------
  {\cal A}_{e}^{SP} &=& \frac{1}{N}
  {\int}_{0}^{1}dx_{1} {\int}_{0}^{1}dx_{2} {\int}_{0}^{1}dx_{3}
  {\int}_{0}^{\infty}b_{1} db_{1} {\int}_{0}^{\infty}b_{2} db_{2}
  {\int}_{0}^{\infty} db_{3}\, {\delta}(b_{2}-b_{3})
   \nonumber \\ & &
  {\phi}_{B_{c}}(x_{2})\, {\phi}_{D_{s}}(x_{3},b_{3})\,
  E_{e}(t_{e})\,{\alpha}_{s}(t_{e})\,
  H_{ef}({\alpha}_{a},{\beta}_{e},b_{1},b_{2})
   \nonumber \\ & &
   \Big\{ {\phi}_{\Upsilon}^{v}(x_{1})\, r_{b}\,(r_{2}+r_{3})
   +{\phi}_{\Upsilon}^{t}(x_{1}) \Big[ r_{2}\,(x_{2}-x_{1})
   +r_{3}\,(\bar{x}_{3}-x_{1}) \Big] \Big\}
   \label{amp-fige-03},
   \end{eqnarray}
  %-----------------------------------------------------
  %-----------------------------------------------------
   \begin{eqnarray}
  {\cal A}_{f}^{LL} &=& \frac{1}{N}
  {\int}_{0}^{1}dx_{1} {\int}_{0}^{1}dx_{2} {\int}_{0}^{1}dx_{3}
  {\int}_{0}^{\infty}b_{1} db_{1} {\int}_{0}^{\infty}b_{2} db_{2}
  {\int}_{0}^{\infty} db_{3}\, {\delta}(b_{2}-b_{3})
   \nonumber \\ & &
  {\phi}_{B_{c}}(x_{2})\, {\phi}_{D_{s}}(x_{3},b_{3})\,
  E_{f}(t_{f})\,{\alpha}_{s}(t_{f})\,
  H_{ef}({\alpha}_{a},{\beta}_{e},b_{1},b_{2})\,
   \Big\{ {\phi}_{\Upsilon}^{v}(x_{1})
   \nonumber \\ & &
   \Big[ \frac{s\,(\bar{x}_{1}-x_{2})}{m_{1}^{2}}
   +2\,r_{3}^{2}\,(x_{3}-x_{1})
   +r_{2}\,r_{3}\,(\bar{x}_{3}-x_{2}) \Big]
   -r_{b}\,{\phi}_{\Upsilon}^{t}(x_{1}) \Big\}
   \label{amp-figf-01}, \\
  %-----------------------------------------------------
  {\cal A}_{f}^{LR} &=& \frac{1}{N}
  {\int}_{0}^{1}dx_{1} {\int}_{0}^{1}dx_{2} {\int}_{0}^{1}dx_{3}
  {\int}_{0}^{\infty}b_{1} db_{1} {\int}_{0}^{\infty}b_{2} db_{2}
  {\int}_{0}^{\infty} db_{3}\, {\delta}(b_{2}-b_{3})
   \nonumber \\ & &
  {\phi}_{B_{c}}(x_{2})\, {\phi}_{D_{s}}(x_{3},b_{3})\,
  E_{f}(t_{f})\,{\alpha}_{s}(t_{f})\,
  H_{ef}({\alpha}_{a},{\beta}_{e},b_{1},b_{2})\,
   \Big\{ {\phi}_{\Upsilon}^{v}(x_{1})
   \nonumber \\ & &
   \Big[ \frac{s\,(x_{1}-x_{3})}{m_{1}^{2}}
   +2\,r_{2}^{2}\,(x_{2}-\bar{x}_{1})
   +r_{2}\,r_{3}\,(\bar{x}_{3}-x_{2}) \Big]
   +r_{b}\,{\phi}_{\Upsilon}^{t}(x_{1}) \Big\}
   \label{amp-figf-02}, \\
  %-----------------------------------------------------
  {\cal A}_{f}^{SP} &=& \frac{1}{N}
  {\int}_{0}^{1}dx_{1} {\int}_{0}^{1}dx_{2} {\int}_{0}^{1}dx_{3}
  {\int}_{0}^{\infty}b_{1} db_{1} {\int}_{0}^{\infty}b_{2} db_{2}
  {\int}_{0}^{\infty} db_{3}\, {\delta}(b_{2}-b_{3})
   \nonumber \\ & &
  {\phi}_{B_{c}}(x_{2})\, {\phi}_{D_{s}}(x_{3},b_{3})\,
  E_{f}(t_{f})\,{\alpha}_{s}(t_{f})\,
  H_{ef}({\alpha}_{a},{\beta}_{e},b_{1},b_{2})
   \nonumber \\ & &
   \Big\{ {\phi}_{\Upsilon}^{v}(x_{1})\, r_{b}\,(r_{2}+r_{3})
   +{\phi}_{\Upsilon}^{t}(x_{1}) \Big[ r_{2}\,(x_{2}-\bar{x}_{1})
   +r_{3}\,(x_{1}-x_{3}) \Big] \Big\}
   \label{amp-figf-03},
   \end{eqnarray}
  %-----------------------------------------------------
  %-----------------------------------------------------
   \begin{eqnarray}
  {\cal A}_{g}^{LL} &=& {\cal A}_{g}^{LR} =
  {\int}_{0}^{1}dx_{2} {\int}_{0}^{1}dx_{3}
  {\int}_{0}^{\infty}b_{2} db_{2}
  {\int}_{0}^{\infty}b_{3} db_{3}\,
  {\phi}_{B_{c}}(x_{2})\, {\phi}_{D_{s}}(x_{3},b_{3})
   \nonumber \\ & &
  E_{f}(t_{g})\, {\alpha}_{s}(t_{g})\,
  H_{gh}({\alpha}_{a},{\beta}_{g},b_{2},b_{3})\,
   \Big\{ x_{2}+r_{3}\,\bar{x}_{2}\, (r_{3}-2\,r_{2}) \Big\}
   \label{amp-figg}, \\
  %-----------------------------------------------------
  {\cal A}_{h}^{LL} &=& {\cal A}_{h}^{LR} =
  {\int}_{0}^{1}dx_{2} {\int}_{0}^{1}dx_{3}
  {\int}_{0}^{\infty}b_{2} db_{2}
  {\int}_{0}^{\infty}b_{3} db_{3}\,
  {\phi}_{B_{c}}(x_{2})\, {\phi}_{D_{s}}(x_{3},b_{3})
   \nonumber \\ & &
  E_{h}(t_{h})\, {\alpha}_{s}(t_{h})\,
  H_{gh}({\alpha}_{a},{\beta}_{h},b_{3},b_{2})\,
   \Big\{ \bar{x}_{3}+r_{2}\,x_{3}\, (r_{2}-2\,r_{3})
   \nonumber \\ & &
   +r_{b}\, (r_{3}-2\,r_{2}) \Big\}
   \label{amp-figh},
   \end{eqnarray}
  %-----------------------------------------------------
  where the mass ratio $r_{i}$ $=$ $m_{i}/m_{1}$;
  $\bar{x}_{i}$ $=$ $1$ $-$ $x_{i}$;
  variable $x_{i}$ is the longitudinal momentum fraction
  of the valence quark;
  $b_{i}$ is the conjugate variable of the
  transverse momentum $k_{i{\perp}}$;
  and ${\alpha}_{s}(t)$ is the QCD coupling at the
  scale of $t$.

  The function $H_{i}$ are defined as follows.
  %-----------------------------------------------------
   \begin{eqnarray}
   H_{ab}({\alpha}_{e},{\beta},b_{i},b_{j})
   &=& K_{0}(\sqrt{-{\alpha}_{e}}b_{i})
   \Big\{ {\theta}(b_{i}-b_{j})
   K_{0}(\sqrt{-{\beta}}b_{i})
   I_{0}(\sqrt{-{\beta}}b_{j})
   + (b_{i}{\leftrightarrow}b_{j}) \Big\}
   \label{hab}, \\
  %-----------------------------------------------------
   H_{cd}({\alpha}_{e},{\beta},b_{2},b_{3}) &=&
   \Big\{ {\theta}(-{\beta}) K_{0}(\sqrt{-{\beta}}b_{3})
  +\frac{{\pi}}{2} {\theta}({\beta}) \Big[
   iJ_{0}(\sqrt{{\beta}}b_{3})
   -Y_{0}(\sqrt{{\beta}}b_{3}) \Big] \Big\}
   \nonumber \\ &{\times}&
   \Big\{ {\theta}(b_{2}-b_{3})
   K_{0}(\sqrt{-{\alpha}_{e}}b_{2})
   I_{0}(\sqrt{-{\alpha}_{e}}b_{3})
   + (b_{2}{\leftrightarrow}b_{3}) \Big\}
   \label{hcd}, \\
  %-----------------------------------------------------
   H_{ef}({\alpha}_{a},{\beta},b_{1},b_{2}) &=&
   \Big\{ {\theta}(-{\beta}) K_{0}(\sqrt{-{\beta}}b_{1})
  +\frac{{\pi}}{2} {\theta}({\beta}) \Big[
   iJ_{0}(\sqrt{{\beta}}b_{1})
   -Y_{0}(\sqrt{{\beta}}b_{1}) \Big] \Big\}
   \nonumber \\ & & \!\!\!\!\!\!\!\!\!\!\!\!\!\!\!\!\!\!\!\!\!\!\!\!
  {\times} \frac{{\pi}}{2} \Big\{ {\theta}(b_{1}-b_{2})
   \Big[ iJ_{0}(\sqrt{{\alpha}_{a}}b_{1})
   -Y_{0}(\sqrt{{\alpha}_{a}}b_{1}) \Big]
   J_{0}(\sqrt{{\alpha}_{a}}b_{2})
   + (b_{1}{\leftrightarrow}b_{2}) \Big\}
   \label{hef}, \\
  %-----------------------------------------------------
  H_{hg}({\alpha}_{a},{\beta},b_{i},b_{j}) &=&
  \frac{{\pi}^{2}}{4}
  \Big\{ iJ_{0}(\sqrt{{\alpha}_{a}}b_{j})
   -Y_{0}(\sqrt{{\alpha}_{a}}b_{j}) \Big\}
   \nonumber \\ &{\times}&
   \Big\{ {\theta}(b_{i}-b_{j})
   \Big[ iJ_{0}(\sqrt{{\beta}}b_{i})
   -Y_{0}(\sqrt{{\beta}}b_{i}) \Big]
   J_{0}(\sqrt{{\beta}}b_{j})
   + (b_{i}{\leftrightarrow}b_{j}) \Big\}
   \label{hgh},
   \end{eqnarray}
  %-----------------------------------------------------
  where $J_{0}$ and $Y_{0}$ ($I_{0}$ and $K_{0}$) are the
  (modified) Bessel function of the first and second kind,
  respectively;
  ${\alpha}_{e}$ (${\alpha}_{a}$) is the
  gluon virtuality of the emission (annihilation)
  diagrams;
  the subscript of the quark virtuality ${\beta}_{i}$
  corresponds to the indices of Fig.\ref{fig1}.
  The definition of the particle virtuality is
  listed as follows.
  %-----------------------------------------------------
   \begin{eqnarray}
  {\alpha}_{e} &=& \bar{x}_{1}^{2}m_{1}^{2}
                +  \bar{x}_{2}^{2}m_{2}^{2}
                -  \bar{x}_{1}\bar{x}_{2}t
   \label{gluon-q2-e}, \\
  %-----------------------------------------------------
  {\alpha}_{a} &=& x_{2}^{2}m_{2}^{2}
                +  \bar{x}_{3}^{2}m_{3}^{2}
                +  x_{2}\bar{x}_{3}s
   \label{gluon-q2-a}, \\
  %-----------------------------------------------------
  {\beta}_{a} &=& m_{1}^{2} - m_{b}^{2}
               +  \bar{x}_{2}^{2}m_{2}^{2}
               -  \bar{x}_{2}t
   \label{beta-fa}, \\
  %-----------------------------------------------------
  {\beta}_{b} &=& m_{2}^{2} - m_{c}^{2}
               +  \bar{x}_{1}^{2}m_{1}^{2}
               -  \bar{x}_{1}t
   \label{beta-fb}, \\
  %-----------------------------------------------------
  {\beta}_{c} &=& x_{1}^{2}m_{1}^{2}
               +  x_{2}^{2}m_{2}^{2}
               +  \bar{x}_{3}^{2}m_{3}^{2}
   \nonumber \\ &-&
                  x_{1}x_{2}t
               -  x_{1}\bar{x}_{3}u
               +  x_{2}\bar{x}_{3}s
   \label{beta-fc}, \\
  %-----------------------------------------------------
  {\beta}_{d} &=& x_{1}^{2}m_{1}^{2}
               +  x_{2}^{2}m_{2}^{2}
               +  x_{3}^{2}m_{3}^{2}
               -  m_{c}^{2}
    \nonumber \\ &-&
                  x_{1}x_{2}t
               -  x_{1}x_{3}u
               +  x_{2}x_{3}s
   \label{beta-fd}, \\
  %-----------------------------------------------------
  {\beta}_{e} &=& x_{1}^{2}m_{1}^{2}
               +  x_{2}^{2}m_{2}^{2}
               + \bar{x}_{3}^{2}m_{3}^{2}
               -  m_{b}^{2}
   \nonumber \\ &-&
                  x_{1}x_{2}t
               -  x_{1}\bar{x}_{3}u
               +  x_{2}\bar{x}_{3}s
   \label{beta-fe}, \\
  %-----------------------------------------------------
  {\beta}_{f} &=& \bar{x}_{1}^{2}m_{1}^{2}
               +  x_{2}^{2}m_{2}^{2}
               +  \bar{x}_{3}^{2}m_{3}^{2}
               -  m_{b}^{2}
   \nonumber \\ &-&
                  \bar{x}_{1}x_{2}t
               -  \bar{x}_{1}\bar{x}_{3}u
               +  x_{2}\bar{x}_{3}s
   \label{beta-ff}, \\
  %-----------------------------------------------------
  {\beta}_{g} &=& x_{2}^{2}m_{2}^{2}
               +  m_{3}^{2}
               +  x_{2}s
   \label{beta-fg}, \\
  %-----------------------------------------------------
  {\beta}_{h} &=& \bar{x}_{3}^{2}m_{3}^{2}
               +  m_{2}^{2}
               +  \bar{x}_{3}s
               - m_{b}^{2}
   \label{beta-fh}.
   \end{eqnarray}
  %-----------------------------------------------------

  The typical scale $t_{i}$ and the Sudakov factor $E_{i}$
  are defined as follows, where the subscript $i$ corresponds
  to the indices of Fig.\ref{fig1}.
  %-----------------------------------------------------
   \begin{eqnarray}
   t_{a(b)} &=& {\max}(\sqrt{-{\alpha}_{e}},\sqrt{-{\beta}_{a(b)}},1/b_{1},1/b_{2})
   \label{tab}, \\
   t_{c(d)} &=& {\max}(\sqrt{-{\alpha}_{e}},\sqrt{{\vert}{\beta}_{c(d)}{\vert}},1/b_{2},1/b_{3})
   \label{tcd}, \\
   t_{e(f)} &=& {\max}(\sqrt{{\alpha}_{a}},\sqrt{{\vert}{\beta}_{e(f)}{\vert}},1/b_{1},1/b_{2})
   \label{tef}, \\
   t_{g(h)} &=& {\max}(\sqrt{{\alpha}_{a}},\sqrt{{\beta}_{g(h)}},1/b_{2},1/b_{3})
   \label{tgh},
   \end{eqnarray}
  %-----------------------------------------------------
  %-----------------------------------------------------
   \begin{equation}
   E_{i}(t) =
   \left\{ \begin{array}{lll}
  {\exp}\{ -S_{{\Upsilon}(1S)}(t)-S_{B_{c}}(t) \}, &~& i=a,b \\
  {\exp}\{ -S_{{\Upsilon}(1S)}(t)-S_{B_{c}}(t)-S_{D_{s}}(t) \}, & & i=c,d,e,f \\
  {\exp}\{ -S_{B_{c}}(t)-S_{D_{s}}(t) \}, & & i=g,h
   \end{array} \right.
   \label{sudakov-exp}
   \end{equation}
  %-----------------------------------------------------
  %-----------------------------------------------------
   \begin{eqnarray}
   S_{{\Upsilon}(1S)}(t) &=&
   s(x_{1},p_{1}^{+},1/b_{1})
  +2{\int}_{1/b_{1}}^{t}\frac{d{\mu}}{\mu}{\gamma}_{q}
   \label{sudakov-bc}, \\
  %-----------------------------------------------------
   S_{B_{c}}(t) &=&
   s(x_{2},p_{2}^{+},1/b_{2})
  +2{\int}_{1/b_{2}}^{t}\frac{d{\mu}}{\mu}{\gamma}_{q}
   \label{sudakov-bc}, \\
  %-----------------------------------------------------
   S_{D_{s}}(t) &=&
   s(x_{3},p_{3}^{+},1/b_{3})
  +2{\int}_{1/b_{3}}^{t}\frac{d{\mu}}{\mu}{\gamma}_{q}
   \label{sudakov-ds},
   \end{eqnarray}
  %-----------------------------------------------------
  where ${\gamma}_{q}$ $=$ $-{\alpha}_{s}/{\pi}$ is the
  quark anomalous dimension;
  the explicit expression of $s(x,Q,1/b)$ can be found in
  the appendix of Ref.\cite{pqcd1}.
  \end{appendix}

   %%%%%%%%%%%%%%%%%%%%%%%%%%%%%%%%%%%%%%%%%%%%%%%%%%%%%%%%%%%
  

\begin{thebibliography}{99}
  \bibitem{pdg}
          K. Olive {\em et al.} (Particle Data Group), Chin. Phys. C 38, 090001 (2014).
  \bibitem{o}
          S. Okubo, Phys. Lett. 5, 165 (1963).
  \bibitem{z}
          G. Zweig, CERN-TH-401, 402, 412 (1964).
  \bibitem{i}
          J. Iizuka, Prog. Theor. Phys. Suppl. 37-38, 21 (1966).
  \bibitem{1212.6552}
          C. Patrignani, T. Pedlar, and J. Rosner, Annu. Rev. Nucl. Part. Sci. 63, 21 (2013).
  \bibitem{pqcd1}
          H. Li, Phys. Rev. D 52, 3958 (1995).
  \bibitem{pqcd2}
          C. Chang,  H. Li, Phys. Rev. D 55, 5577 (1997).
  \bibitem{pqcd3}
          T. Yeh, H. Li, Phys. Rev. D 56, 1615 (1997).
  \bibitem{1212.5342}
          J. Brodzicka {\em et al.} (Belle Collaboration), Prog. Theor. Exp. Phys. 2012, 04D001.
  \bibitem{zpc62.271}
          M. Sanchis-Lozano, Z. Phys. C 62, 271 (1994).
  \bibitem{qcdf2}
          M. Beneke {\em et al.}, Nucl. Phys. B 591, 313 (2000).
  \bibitem{scet1}
          C. Bauer {\em et al.},  Phys. Rev. D 63, 114020 (2001).
  \bibitem{scet2}
          C. Bauer, D. Pirjol, I. Stewart, Phys. Rev. D 65, 054022 (2002).
  \bibitem{scet3}
          C. Bauer {\em et al.}, Phys. Rev. D 66, 014017 (2002).
  \bibitem{scet4}
          M. Beneke {\em et al.}, Nucl. Phys. B 643, 431 (2002).
  \bibitem{9512380}
          G. Buchalla, A. Buras, M. Lautenbacher, Rev. Mod. Phys. 68, 1125, (1996).
  \bibitem{npb366}
          S. Catani, M. Ciafaloni, and F. Hautmann, Nucl. Phys. B 366, 135 (1991).
  \bibitem{prd22}
          G. Lepage, S. Brodsky, Phys. Rev. D 22, 2157 (1980).
  \bibitem{prd78lv}
          R. Li, C. L\"{u}, H. Zou, Phys. Rev. D 78, 014018 (2008).
  \bibitem{prd46}
          G. Lepage {\em et al.}, Phys. Rev. D 46, 4052 (1992).
  \bibitem{prd51}
          G. Bodwin, E. Braaten, G. Lepage, Phys. Rev. D 51, 1125 (1995).
  \bibitem{rmp77}
          N. Brambilla {\em et al.}, Rev. Mod. Phys. 77, 1423 (2005).
  \bibitem{prd92}
          J. Sun {\em et al.}, Phys. Rev. D 92, 074028 (2015).
  \bibitem{fbc}
          T. Chiu, T. Hsieh, C. Huang, K. Ogawa, Phys. Lett. B 651, 171 (2007).
  \bibitem{jhep1407}
          R. Aaij {\em et al.} (LHCb Collaboration), JHEP 1407, 094 (2014).
  \bibitem{plb740}
          B. Abelev {\em et al.} (ALICE Collaboration), Phys. Lett. B 740, 105 (2015).
  \end{thebibliography}
  \end{document}